\documentclass[Royal,sagev]{sagej}   

\usepackage{amsthm,amssymb,amsmath,amsxtra,array,amsfonts,dsfont,enumerate,graphicx,multirow,moreverb,natbib,setspace,subfig}

\usepackage[colorlinks,bookmarksopen,bookmarksnumbered,citecolor=black,urlcolor=black]{hyperref}

\usepackage{rotating}

\theoremstyle{plain}

\theoremstyle{definition}

\theoremstyle{remark}

\begin{document}

\runninghead{Yun Li \textsc{et al}}

\title{Using Multiple Imputation to Classify Potential Outcomes Subgroups}

\author{Yun Li\affilnum{1,2,3}, Irina Bondarenko\affilnum{3}, Michael R. Elliott\affilnum{3}, Timothy P. Hofer\affilnum{4,5} and Jeremy M.G. Taylor\affilnum{3}  }

\affiliation{\affilnum{1} Division of Biostatistics, University of Pennsylvania, Philadelphia, PA.\\
\affilnum{2} Department of Pediatrics, Children's Hospital of Philadelphia, Philadelphia, PA.\\
\affilnum{3} Department of Biostatistics, University of Michigan, Ann Arbor, MI. \\
\affilnum{4} Division of General Medicine, University of Michigan, Ann Arbor, MI. \\
\affilnum{5} VA Health Service Research \& Development Center for Clinical Management Research, Ann Arbor, MI.}

\corrauth{Yun Li}

\email{yun.li@pennmedicine.upenn.edu}

\begin{abstract}
With medical tests becoming increasingly available, concerns about over-testing, over-treatment and health care cost dramatically increase. Hence, it is important to understand the influence of testing on treatment selection in general practice. Most statistical methods focus on average effects of testing on treatment decisions. However, this may be ill-advised, particularly for patient subgroups that tend not to benefit from such tests. Furthermore, missing data are common, representing large and often unaddressed threats to the validity of most statistical methods. Finally, it is often desirable to conduct analyses that can be interpreted causally. Using the Rubin Causal Model framework, we propose to classify patients into four potential outcomes subgroups, defined by whether or not a patient's treatment selection is changed by the test result and by the direction of how the test result changes treatment selection. This subgroup classification naturally captures the differential influence of medical testing on treatment selections for different patients, which can suggest targets to improve the utilization of medical tests. We can then examine patient characteristics associated with patient potential outcomes subgroup memberships. We used multiple imputation methods to simultaneously impute the missing potential outcomes as well as regular missing values. This approach can also provide estimates of many traditional causal quantities of interest. We find that explicitly incorporating causal inference assumptions into the multiple imputation process can improve the precision for some causal estimates of interest. We also find that bias can occur when the potential outcomes conditional independence assumption is violated; sensitivity analyses are proposed to assess the impact of this violation. We applied the proposed methods to examine the influence of 21-gene assay, the most commonly used genomic test in the United States, on chemotherapy selection among breast cancer patients.
\end{abstract}


\maketitle

\section{Introduction}\label{sec1}

Newly diagnosed cancer patients and their physicians frequently face the challenges of deciding which tests and subsequent treatments to use. The testing itself can be both invasive and expensive, and the cost of medical tests reaches 70 billion dollars annually \cite{Brill}. Furthermore, test results do not predictably inform treatment decisions, either because guidelines are not always available or they are not followed in practice. Thus, describing how testing impacts the choice of treatment in general practice is of fundamental importance. We will view this problem in a causal framework, in which taking the test is the exposure of interest, and the treatment received is the outcome of interest.

An example is the use of the 21-gene recurrence score (RS) assay in breast cancer care, the most commonly used genomic assay in the United States. The test is used to inform recommendations about the use of chemotherapy. Women who get a high score on the test are recommended to take chemotherapy to reduce their risk of recurrence, whereas women with a low score are not because they have a good prognosis and are expected to receive little additional benefit from chemotherapy. Prior studies have demonstrated that RS tests can provide more precise estimation of the risk of cancer recurrence and the need for chemotherapy than do traditional histopathologic features (e.g., tumor grade and size) \cite{Albain, Schnipper, Paik1, Habel, Paik2, Dowsett, Goldstein}. Even though the current clinical guidelines only endorse the use of RS assay in selected patients with certain clinical characteristics \cite{Harris, NCCN}, more and more patients who do not meet these criteria are being selected for RS testing. Also, chemotherapy is highly morbid, and many believe it is being over-prescribed \cite{Sparano, Manj}. Thus, there is a pressing need to measure the influence of the RS assay on chemotherapy decisions to evaluate whether the test is being used in an efficient manner and acted upon appropriately in general practice.

Determining the influence of testing on treatment selection in a community practice setting is not well studied and is challenging to quantify using standard methods. First, selection bias may occur because testing is not offered randomly. Both clinical factors as well as patient and physician preferences influence whether testing is offered, and this confounds the effect of testing on treatment selection \cite{Bosco, Salas}. Second, while causal inference methods such as propensity-score based methods can address selection bias and give effect estimates with desirable causal interpretations, they are limited to estimating an overall effect of testing on treatment selection, not how testing changes treatment plans within patient subgroups. The overall effect can obscure the degree to which treatment plans are changed by testing. For example, the overall effect can be zero, if testing encourages and discourages treatment for an equal number of patients, even though testing did change treatment plans. Third, missing data are common, especially when multiple data sources are combined as in many studies. Missing data are often-unaddressed threats to the validity of statistical methods.

In this paper, we propose to use a potential outcomes framework (i.e., a causal inference framework) \cite{Holland} to measure the influence of testing by comparing treatment use when everyone is tested with when no one is tested. This minimizes selection bias affecting test use that is attributable to observed confounders. Specifically, we propose to classify patients into four potential outcomes subgroups defined by their expected treatment decisions with a given test versus without the test: 1) Never-Treated: those who would not have been treated with or without the test; 2) Treatment-Encouraged: those who would have been treated with the test but would not have been treated without the test; 3) Treatment-Discouraged: those who would not have been treated with the test but would have been treated without the test; 4) Always-Treated: those who would have been treated with or without the test. These four subgroups (listed in Table 1) are called potential outcomes subgroups \cite{Holland}. A patient’s membership in these subgroups may be driven by patient characteristics or by systemic factors (e.g., treating physician). Such classification can also suggest ways to improve the utilization of medical tests in treatment decisions. The classification recognizes that a medical test can have different influences on treatment decisions for different patients; i.e., there may be effect heterogeneity of testing on treatment. With the classification, we will be able to measure how many patients’ treatment selections are changed (or unchanged) by test results for overall and key clinical subgroups, and also quantify how much test results encourage (or discourage) treatment selections. The potential outcomes subgroups bear some resemblance to principal stratification \cite{Frangakis}, which has been extensively used by us and others to study the causal effect of intermediate variables (e.g., degree of compliance, surrogate markers) \cite{Huang, Li, Roy}. However, unlike principal strata, which are defined by the joint distribution of intermediate variables under different exposure status, here we focus on the joint distribution of the outcome of interest (here treatment received) under different exposure status (here test received). We believe this framework provides important conceptual and practical advances in studying the influence of medical testing in treatment selection. We will also demonstrate its advantages of providing unique and highly relevant causal quantities of interest over alternative causal inference methods.

We will classify patients into these potential outcomes subgroups by exploring the close connection between missing data analysis and causal inference. Since two potential outcomes of the same patients are not simultaneously observed, we will use multiple imputation methods to impute the unobserved potential outcomes to enable us to make causal inferences. Multiple imputation methods have been used to impute missing potential outcomes in solving different problems in causal inference literature \cite{Rubin, Taylor, Gutman, Bondarenko, Westreich, Miao, Zhou}. These methods have the advantage of simultaneously handling standard missing data in observable variables (e.g. due to non-response) and missing potential outcomes. However, since missing data analysis and causal inference have different objectives and data structures, there are important distinctions between them. First, causal inference focuses on comparing the potential outcomes of the same patient under different exposure levels; while missing data analyses do not. Second, the causal assumptions (e.g., ignorability of treatment assignment or potential outcomes conditional independence) are not usually made in standard missing data analyses. Third, causal inference distinguishes pre-exposure variables from post-exposure variables and usually does not use a post-exposure variable, while missing data analyses do not differentiate them and often utilize a post-exposure variable in the imputation models if it is predictive of other variables with missing values. Fourth, the missing potential outcomes problem is highly structured such that only one of two potential outcomes is observed for any patient when the exposure variable is binary. In this situation, the missing data indicator is identical to the binary exposure variable. Fifth, since no two potential outcomes of the same patient can be observed simultaneously, there is no information about the correlation between two potential outcomes of the same patient. This distinct data structure is similar to the missing data structure in the literature on "statistical matching" \cite{Kim, Conti}. More detailed discussions about missing data analyses and causal inference can be found by Ding and Li (2018) and Westreich et al (2015) \cite{Ding, Westreich}. In  this  paper,  we will study the performance of our proposed methods. We will also examine the efficiency gain of our causal estimates when we incorporate the knowledge of our data  structure  and  causal  inference  assumptions  into  the  multiple  imputation process. We will propose sensitivity analyses to investigate the violation of one of the key causal inference assumptions on causal estimates (i.e., potential outcomes conditional independence). Additionally, we will explore the impact of incorporating a post-exposure variable in the multiple imputation process on the causal estimates. Furthermore, we will demonstrate that this approach is more robust towards certain types of model mis-specification than other causal methods (e.g., propensity-score based methods).

\begin{table}
\begin{center}
\caption{Potential outcomes subgroups defined by the effect of testing on treatment selection.}
\begin{tabular}{lll}
\hline
&  \multicolumn{2}{l}{Treatment Selection with Test ($Y_1$)} \\
 \hline
Treatment Selection & & \\
without Test ($Y_0$) & Yes & No \\
\hline
 Yes & Always Treated & Treatment Discouraged \\
 No & Treatment Encouraged & Never Treated \\
\hline
\end{tabular}
\end{center}
\label{Tab:1}
\end{table}

To illustrate our methods, we will examine the impact of the RS assay on the decision to undergo chemotherapy for breast cancer patients. The analyses will use combined data sets from the Surveillance, Epidemiology, and End Results databases (SEER) registry, genomic testing laboratories, and patient/physician surveys, recently collected from the ICanCare Study \cite{Friese, LiY}. The data linkage was a unique survey-registry-industry collaboration and the ICanCare study was part of the only program project funded by National Cancer Institute that focused on cancer treatment decision making.

\section{Notation, the Causal Framework and Quantities of Interest}\label{sec2}
Let $Y$ represent the treatment receipt (i.e., the outcome of interest, 0/1 for no/yes), $Z$ the status of being tested (i.e., the exposure of interest, 0/1 for no/yes), $X$ a set of covariates and $R$ the test result. Naturally, $R$ only exists for testers and is a post-exposure variable. We assume some observations are missing in $Y$, $R$ and in some variables of $X$. Let $M$ be the associated missing-data indicator matrix for these observable variables. The observed and missing parts of $Y$ are denoted by $Y^{obs}$ and $Y^{mis}$ respectively, and the same for $X$.

For each patient, we assume there are two potential outcomes, $Y_1$ and $Y_0$, corresponding to being tested or not ($Z=1,0$). Let $Y_1$ indicate a patient’s treatment status that would be observed if the patient is tested and $Y_0$ otherwise. If we know both $Y_1$ and $Y_0$ for every patient, we can infer whether a test causes a change in treatment selection by comparing $Y_1$ to $Y_0$, and calculate many causal quantities that are free of selection bias. In this manuscript, we choose several representative causal quantities including: (1) average testing effect (ATE), E($Y_1-Y_0$), for the overall population and patient subgroups; (2) marginal odds ratio (MOR), $[P_1(1 - P_0)]/[P_0 (1 - P_1)]$ where $P_0 = P(Y_0=1)$ and $P_1 = P(Y_1 = 1)$; and (3) the fractions of patients in each of  the four potential outcome subgroups: ($Y_0=0$, $Y_1=0$) for the never-treated, ($Y_0=0$, $Y_1=1$) for the treatment-encouraged, ($Y_0=1$, $Y_1=0$) for the treatment-discouraged, and ($Y_0 =1$, $Y_1 =1$) for the always-treated group, which are denoted by $P(00)$, $P(01)$, $P(10)$ and $P(11)$ respectively. The MOR defined in (2) compares the odds of treatment use if everyone were tested to the odds if no one were tested. Odds ratios in observational studies are usually association measures and lack causal interpretations because of confounding. In contrast, MORs do have causal interpretations. Based on the fractions defined in (3), we can calculate many quantities that are functions of them. For example, we can calculate the proportion of patients whose treatment plans are changed by the use of testing (i.e., the proportion of patients who were either treatment encouraged or discouraged), $P(01) + P(10)$, and the proportion of patients whose treatment plans are not changed (i.e., the proportion of patients who were either never-treated or always-treated), $P(00)+P(11)$. In fact, ATE is also a function of these fractions, which is equal to the difference in the proportions in the ``treatment encouraged'' and ``treatment discouraged'' subgroups, i.e., $P(01) - P(10)$. We are also interested in estimating the associations between patient characteristics and patient membership in these potential outcomes subgroups.

\section{Estimating the Distributions of Potential Treatment Selection}
To measure the causal influence of a test, ideally, we wish to know the treatment plan for each patient when she is tested and when she is not. That is, we wish to know both $Y_0$ and $Y_1$. However, the treatment plan is observable for a given patient in only one of these two scenarios because she may not get tested and; even if she does, her pre-test treatment plan is not usually documented. Formally, this is a missing data problem. We use sequential regression multiple imputation (or sometimes called the fully conditional specification method) \cite{Raghu, Van} to impute the unobserved (counterfactual) treatment plan for each patient, which we term the potential outcomes multiple imputation method (POMI). That is, for patients who received the test, we impute what would have been their treatment plans had they not received the test. And for patients who did not receive the test, we impute what would have been their treatment plans had they received the test from the predictive distribution of $Y$ given $X$ and/or $R$. Once we have complete information for $Y_0$ and $Y_1$ after imputation, we can calculate the causal quantities of interest. Since we have incomplete information in observable variables $Y$, $X$ and $R$, we can simultaneously impute missing potential outcomes and missing values in these observable variables during the multiple imputation process using the same imputation methods.

Although we can view causal inference as a missing data problem, there are distinct features of causal inference different from standard missing data problems. As stated in the Introduction section, the objectives of causal inference are to obtain the causal effect of an exposure on the outcome, while missing data analyses are to handle missing values in variables, utilize incomplete information and reduce bias due to non-response. Hence, some assumptions are closely related but others are different. Below we list four assumptions used in missing data and/or causal inference literature:
\begin{enumerate}
    \item Missing at random \cite{Little}, $f(M|Z, Y, X, R) = f(M|Z, Y^{obs}, X^{obs}, R^{obs})$. That is, missingness can be explained by the observed data. This is a standard assumption in the missing data literature. Note that $f()$ denotes a probability density or mass function for the distribution of either a continuous or discrete variable in this manuscript.
    \item Stable unit treatment value assumption (SUTVA) \cite{Rubin, Rubin1}. The potential outcomes of one patient are unaffected by the potential outcomes or testing status of other patients. This is a standard assumption in the causal inference literature. This assumption is not usually explicitly made, but it is implied in the missing data literature.
    \item Ignorability of exposure \cite{Rubin2}, $(Y_0, Y_1)\bot Z | X$. This assumption means that the receipt of test is ``randomized'' for subjects with the same set of covariates $X$, that is, there are no unmeasured confounders between $Z$ and $Y$ after adjustment for $X$. It is a standard assumption in the causal inference literature, and a special case of the missing-at-random assumption (see Assumption 1).
    \item \textbf{Potential outcomes conditional independence}, $(Y_0 \bot Y_1)| X$. This assumption states that two potential outcomes are conditionally independent. However, since the two potential outcomes are never observed together, we have no information about the correlation, independence or joint distribution of $Y_0$ and $Y_1$ from the observed data. This assumption is frequently made in the causal inference literature \cite{Zhang, Efron, Daniels}. It is also made in the missing data and survey methodology literature when applying the less-known statistical matching technique to conduct a joint analysis on variables that are never jointly observed \cite{Kim, Conti}. We will explore robustness of our methods to this conditional independence assumption using sensitivity analyses.
\end{enumerate}
These assumptions will be essential to facilitating identification and estimation of causal quantities of interest.

We propose to impute the missing values in $X$ and $R$ as well as the missing potential outcomes ($Y_0$ and $Y_1$) iteratively variable-by-variable using a sequence of conditional regression models until convergence: $f(X_{j} | X_{(-j)}, Z, Y_0, Y_1, R)$, $f(R|X, Z, Y_0, Y_1)$, $f(Y_0| Y_1, Z, X, R)$, $f(Y_1| Y_0, Z, X, R)$ for any $j$ \cite{Kenn, Van, Raghu}. Note that $X = (X_{1}, \cdots, X_{k})$ and $X_{(-j)} = (X_{1}, \cdots, X_{(j-1)}, X_{(j+1)}, \cdots, X_{k})$ denoting the collection of the $k-1$ variables in $X$ except $X_{j}$. This method is the sequential regression multiple imputation method and has its advantages in handling non-normal data \cite{Raghu}. We can evaluate how well the missing variables are imputed as in regular missing data imputation problems.

We impute the missing values $D$ times to give a range for each imputed value to reflect prediction uncertainty. We then obtain the estimates of the quantities of interest for each of the $D$ imputed data sets before combining these estimates and their variances across $D$ imputed data sets using Rubin's formula \cite{Rubin2}. Let $\Delta$ represent any quantity of interest, $\widehat{\Delta}_d$ the point estimate and $W_d$ the within-imputation variance estimate of $\widehat{\Delta}_d$ for the $d$th imputed data set. The POMI estimate of $\Delta$ is obtained by averaging across $D$ imputed data sets, such that $\overline{\Delta} = \frac{1}{D} \sum_{d=1}^D \widehat{\Delta}_d$ where $d=1, \cdots, D$. The MI estimate of the variance of $\overline{\Delta}$ is $V = \overline{W} + (1 + \frac{1}{D})B$, where $\overline{W} = \sum_{d=1}^D W_d/D$, $B = \sum_{d=1}^D (\widehat{\Delta}_d - \overline{\Delta})^2 / (D-1)$. Note that $\overline{W}$ represents the average within-imputation variance and $B$ represents the between-imputation variance. The estimated $\Delta$ follows a $t$ distribution, with the degree of freedom
$(D-1)[1+\frac{D}{D+1}(\overline{W}/B)^2]$ \cite{Schenker}.

Specifically, the within-imputation variance for the estimated causal quantity of interest, ATE, for each imputed data set is calculated as follows: $\frac{1}{N-1} \left\{\left[\widehat{P}_{0}(1- \widehat{P}_{0})\right] + \left[\widehat{P}_{1}(1- \widehat{P}_{1})\right] - 2(\widehat{P}_{11} - \widehat{P}_{0}\widehat{P}_{1})\right\}$, where $P_{11} = P(Y_0=1, Y_1=1)$ and $N$ is the sample size. The calculation of the within-imputation variance of the MOR estimate is more complicated than ATE because it is a non-linear function of $P_0$ and $P_1$. Since MOR is skewed and positive, we calculate the variance for the MOR estimate on a log scale. Specifically, for each imputed data set, we calculate the variance of $log(\widehat{MOR})$ using the delta method as follows: $\left[\widehat{P}_{0} + \widehat{P}_{1} - (\widehat{P}_{0} - \widehat{P}_{1})^2 - 2\widehat{P}_{11} \right] / \left[N \widehat{P}_{0} \widehat{P}_{1} ( 1 - \widehat{P}_{0})(1 - \widehat{P}_{1})\right]$. From the variance of $log(\widehat{MOR})$, we can then calculate the variance of the estimated MOR using the delta method again. The details for the variance calculation of estimated MOR are provided in the appendix. The within-imputation variances for estimated $P(00)$, $P(01)$, $P(11)$ and $P(10)$ are straightforward to calculate as multinomial proportions.

What we described so far are standard multiple imputation procedures. We will explore several features of causal inference that are distinct from standard missing data problems and see how the incorporation of them in the multiple imputation process impacts the causal estimates. These features will be incorporated through the four different POMI methods investigated in the following Simulation section. First, we will explicitly incorporate the causal inference assumptions (listed above as Assumptions 2, 3 and 4) and distinct data structures into the multiple imputations of missing potential outcomes. Under Assumption 2 and 3, we can simplify our imputation models for $Y_0$ and $Y_1$ such that $f(Y_0|Y_1, Z, X, R) = f(Y_0|Y_1, X, R)$ and $f(Y_1|Y_0, Z, X, R) = f(Y_1| Y_0, X, R)$. Under Assumptions 2, 3 and 4, we can further simplify our imputation models for $Y_0$ and $Y_1$ such that $f(Y_0|Y_1, Z, X, R) = f(Y_0|X, R)$ and $f(Y_1|Y_0, Z, X, R) = f(Y_1| X, R)$. Considering $R$ only exists for testers, we can further simplify the models such that $f(Y_0|Y_1, Z, X, R) = f(Y_0|X)$. Second, we will explore the use of $Z$ in the multiple imputation process. $Z$ is a missing data indicator for $Y_0$ and $Y_1$. With an MAR assumption, multiple imputation does not need to include the missing data indicator \cite{Van}. However, here $Z$ also serves the dual role as the exposure of interest. Third, we will investigate the use of test results $R$, a correlate of $Y$ and a post-exposure variable. In causal inference, post-exposure variables are not usually utilized because they may be in the causal pathway, and adjusting for them can bias results \cite{Robins}. However, in standard missing data problems, post-exposure variables are usually included in the imputation models to increase the predictibility of variables with missing values, when post-exposure variables are correlated with the variables with missing values and hence contain additional information about them.

\section{Simulations}
We conduct simulations to examine: 1) the impact of $Z$, $R$ and causal inference assumptions on estimating the quantities of interest through comparisons of four POMI methods, which differ by whether the imputation models include $Z$, and/or $R$ or make certain causal inference assumptions; 2) the properties of causal estimates of ATE and MOR estimated by these four POMI methods and the inverse probability weighting method (IPW); 3) the properties of the unique causal estimates that are obtainable from the four POMI methods; 4) the robustness of the POMI methods towards model misspecification in comparisons with IPW; 5) sensitivity analyses towards the potential outcomes conditional independence assumption (Assumption 4); 6) the properties of the estimated association between patient characteristics and potential outcomes subgroup memberships through the POMI methods.

\subsection{Simulation Design}
We assume that $X$ includes two variables, $X_1$ and $X_2$. The distributions of $X_1$, $X_2$ and test results $R$ follow a multivariate normal distribution of
\begin{eqnarray}
\left( \begin{array}{c} X_1 \\ X_2 \\ R^*
\end{array} \right) \sim MVN \left( \left(\begin{array}{c} 0 \\ 1 \\ -1
\end{array} \right), \left(\begin{array}{ccc} 1 &
0.3  & 0.3 \\  0.3 &
1 & 0.3 \\ 0.3 &
0.3 & 1 \end{array} \right) \right). \label{invar}
\end{eqnarray}
We further categorize $R^*$ into three groups to mimic the distribution of RS assay results in our data example. We let $R = 1$ if $R^*<=-0.747$, $R = 2$ if $-0.747<R^*<=0.282$ and $R=3$ if $R^*>0.282$. The distributions of $Z$, $Y_0$ and $Y_1$ are specified as follows: $\textit{logit}[P(Z=1)] = -1.5 - X_1 + 0.5 X_2+ \alpha X_1X_2$, $\textit{logit}[P(Y_0=1)] = 0.5 X_1 + X_2 + \beta_1 U$ and $\textit{logit}[P(Y_1=1)] = 0.5 X_1 + X_2  - 1.5 I(R_1=1) - 0.5 I(R_1=2) + I(R_1=3) + \beta_2 U$. Note that $U$ is an unmeasured predictor of $Y$, which is used to induce a non-zero correlation between $Y_0$ and $Y_1$ after adjusting for $X$ and hence a violation of the potential outcomes conditional independence assumption. We assume $U \sim$ Normal(0,1). The default model parameter specifications in our simulations are: $\alpha =0$ (i.e., no interaction between $X_1$ and $X_2$), and $\beta_1=\beta_2=0$ (i.e., conditional independence between potential outcomes). However, we will let $\alpha \neq 0$  when conducting analyses to test the robustness of our methods towards model misspecifications. We will also allow $\beta_1$ and $\beta_2$ to be non-zero when conducting sensitivity analyses on the potential outcomes conditional independence assumption (Assumption 4).

Each simulated data set consists of $X_1$, $X_2$, $Z$, $R$ and $Y$. To be consistent with reality, the data set used in the data example does not have information on $U$, or on $R$ when $Z=0$. Further, we allow $X_2$ and $Y$ be partially missing. Let $M_x$ and $M_y$ be $0$ or $1$, indicating $X_2$ and $Y$ being available or missing, respectively. The models for the probability of missing $X_2$ and $Y$ are specified as follows: $\textrm{logit}[P(M_x=1)] = -1 + 0.2X_1$ and $\textrm{logit}[P(M_y=1)] = -2.5 + X_1$. We assume $X_1$ is completely measured. For each parameter specification, we simulate $500$ data sets, each of which has $N=1,000$ subjects. Unless specified otherwise, the default scenarios in our data simulations satisfy Assumptions 1-4. Additional simulations with $N=500$ and $250$ subjects are also conducted to assess the impact of different sample sizes and summarized in Appendix.

We investigate the performance of four POMI methods. These imputation methods differ by whether the imputation models include $Z$, $R$ or explicitly make causal assumptions of ignorabilty of exposure and potential outcomes independence (Assumptions 3 and 4). We will investigate the impact of these choices on estimates of the causal quantities of interest. These choices are reflected by different conditional regression models as described below. All four methods make the MAR and SUTVA assumptions (Assumptions 1 and 2), and are carried out by the IVEWARE software \cite{Raghu}.
\begin{enumerate}
    \item Potential outcome imputation method 1 (POMI-Z): this is the default imputation procedure carried out by IVEWARE which consists of the following sequence of conditional regression models until convergence: $f(X_2|X_1, Y_0, Y_1, R)$, $f(R|X_1, X_2, Y_0, Y_1)$, $f(Y_0|X_1, X_2, Y_1, R)$, and $f(Y_1| X_1, X_2, Y_0, R)$. Note that this imputation method does not make Assumptions 3 and 4, or utilize $Z$ (the exposure variable and the missing indicator for $Y_0$ and $Y_1$), but it utilizes the post-exposure variable $R$.
    \item Potential outcome imputation method 2 (POMI): different from POMI-Z, we now include $Z$ in the imputation of missing $X_2$, using the following sequential regression models: $f(X_2|X_1, Y_0, Y_1, R, Z)$, $f(R|X_1, X_2, Y_0, Y_1)$, $f(Y_0|X_1, X_2, Y_1, R)$, and $f(Y_1| X_1, X_2, Y_0, R)$. The rationale of using the test status $Z$ in imputing missing $X_2$ is that, the probability of receiving the test $Z$ is associated with $X_2$; hence, $Z$ contains information about $X_2$. Similar to POMI-Z, this method does not make Assumptions 3 and 4, but it utilizes $R$.
    \item Potential outcome imputation method 3 (POMI+IND): different from POMI, we now incorporate Assumptions 3 and 4, and consider the fact that test results $R$ are only available among testers. The sequential regression models are simplified as follows: $f(X_2|X_1, Y_0, Y_1, R, Z)$, $f(R| X_1, Y_1, X_2)$, $f(Y_0|X_1, X_2)$, and $f(Y_1|X_1, X_2, R)$. This is also consistent with the default setup in our simulations.
    \item Potential outcome imputation method 4 (POMI+IND-R): different from POMI+IND, we now exclude the post-exposure variable $R$, using the following sequential regression models: $f(X_2|X_1, Y_0, Y_1, Z)$, $f(Y_0|X_1, X_2)$, and $f(Y_1| X_1, X_2)$.
\end{enumerate}
Note that $f(X_2|.)$ represents a conditional probability density function modelled by a linear regression for $X_2$, $f(R|.)$ represents a conditional probability mass function modelled by a multinomial regression for $R$, $f(Y_0|.)$ and $f(Y_1|.)$ represent conditional probability mass functions modelled by logistic regression for $Y_0$ and $Y_1$ respectively. We iteratively impute the missing values in $X_2$, $R$, $Y_0$ and $Y_1$, with the sequential conditional regression models until convergence. To evaluate how well the missing variables are imputed, we compare the distributions of observed and imputed values conditional on the estimated propensity of being observed \cite{Bond}.  We repeat the imputation process 5 times to have 5 imputed data sets. We then conduct statistical inference on these 5 imputed data sets.

We compared these four methods with the inverse probability weighting method (IPW) \cite{Horvitz} which is a propensity-score based method and commonly used to obtain causal inference. Before using IPW, we need to address the issue of missing data first. We use the same sequential regression multiple imputation method to impute missing values in the observable variables $X_2$ and $Y$ 5 times with the following sequence of conditional regression models: $f(X_2|X_1, Y, Z)$, $f(Y|X_1, X_2, Z)$. For each of the 5 imputed data sets, we calculate the propensity score (denoted by $\pi_i = P(Z_i=1)$) from a logistic regression of $Z$ as a function of $X_1$ and $X_2$. We then estimate  $P(Y_1=1)$ and $P(Y_0=1)$ as $\widehat{P}_{1, ipw} = \widehat{P}_{ipw}(Y_1=1) = \left(\sum_i^N{\frac{Z_i Y_i}{\widehat{\pi}_i}}\right)/N$ and $\widehat{P}_{0, ipw} = \widehat{P}_{ipw}(Y_0=1) = \left[\sum_i^N{\frac{(1 - Z_i)Y_i}{1 - \widehat{\pi}_i}}\right]/N $, respectively. Hence, the estimated ATE is $\widehat{ATE}_{IPW} = \widehat{P}_{1, ipw} - \widehat{P}_{0, ipw}$. The marginal odds ratio is estimated as $\widehat{MOR}_{IPW} = \left[\widehat{P}_{1, ipw}(1 - \widehat{P}_{0, ipw})\right]/ \left[\widehat{P}_{0, ipw}(1 - \widehat{P}_{1, ipw})\right]$. The variance calculations of the estimated ATE and MOR use the delta method. For MOR, the log transformation is used to obtain the variance of $log(\widehat{MOR})$ first before obtaining the variance of $\widehat{MOR}$, similar to what has been described in the Appendix. Finally, Rubin's formulae are used to combine the estimates and their variances across 5 imputed data sets.

The true values for the causal quantities of interest cannot be obtained analytically but can be closely approximated by simulating $10$ million patients with complete information on both $Y_0$ and $Y_1$, based on which the true causal quantities can be obtained. This approach has often been used in causal inference research. For each set of parameter specifications, we simulate $500$ data sets. For each quantity of interest, we estimate the mean of each estimate, bias (the difference between the estimate and the true value), coverage probability (CR) of the $95\%$ confidence interval, standard error (SE), empirical standard deviation (ESD) and mean squared error (MSE) across $500$ data sets.

\subsection{Simulation Results}

First, we evaluate the properties of estimates for ATE and MOR from these five different methods (Table 2). These quantities of interest are obtainable from all methods. Based on the evaluation, POMI-Z has larger bias and lower coverage rates while all other methods give negligible bias and close to nominal $95\%$ coverage rates. The IPW method tends to give slightly bigger variance estimates; while all the POMI methods give similar variance estimates. Additional simulations with $N=500$ and $250$ show that the bias of the estimates becomes larger as the sample size decreases in most scenarios; however, they demonstrate similar patterns across different methods as N=1000 (See Tables 8-11 in Appendix).

\begin{table}
\begin{center}
\caption{Comparison of five methods in estimating average testing effect (ATE), and marginal odds ratios (MOR). ESD: empirical standard deviation; SE: average standard error; CR: coverage rate of 95$\%$ confidence interval. Here we assume potential outcomes conditional independence and no interaction between $X_1$ and $X_2$ (i.e., $\alpha=\beta_1=\beta_2=0$).}
\begin{tabular}{llccccl}
\hline
Quantities & Methods & True & 100 $\times$ Bias & ESD & SE & 100 $\times$ CR \\
\hline
ATE & POMI-Z & -0.186 & 1.8 & 0.032 & 0.033 & 90.6 \\
  & POMI & -0.186 & 0.2 & 0.033 & 0.033 & 94.6 \\
  & POMI+IND & -0.186 & 0.0 & 0.032 & 0.033  & 94.8 \\
  & POMI+IND-R & -0.186 & -0.0 & 0.032 & 0.033  & 95.2 \\
  & IPW & -0.186 & 0.9 & 0.035 & 0.039  & 96.6 \\
\hline
MOR & POMI-Z & 0.458 & 4.1 & 0.070 & 0.071 & 90.8 \\
  & POMI & 0.458 & 0.7 & 0.067 & 0.068 & 94.0 \\
  & POMI+IND & 0.458 & 0.3 & 0.065 & 0.067 & 94.4 \\
  & POMI+IND-R & 0.458 & 0.3 & 0.066 & 0.067 & 94.6 \\
  & IPW & 0.458 & 0.0 & 0.066 & 0.074 & 95.8 \\
\hline
\end{tabular}
\end{center}
\label{Tab:1}
\end{table}

Second, we evaluate the properties of the estimated proportions of patients in each potential outcome subgroup by these five methods (Table 3). We estimate the fraction of patients that are either never-treated $P(00)$, treatment-encouraged $P(01)$, treatment-discouraged $P(10)$, or always-treated $P(11)$. Note that IPW cannot provide estimates of these proportions; hence, no estimates from IPW are provided. The simulation results indicate that POMI+IND and POMI+IND-R give estimates with smaller bias, smaller ESD and SE than POMI-Z and POMI. The coverage rates from POMI+IND and POMI+IND-R are also closer to the nominal $95\%$ coverage rates. Note that the conservative coverage rates from POMI-Z and POMI result from the variance being severely overestimated, which overwhelms the bias in these methods. Hence, the results demonstrate the benefit of taking advantage of the causal assumptions of ignorability and potential outcomes conditional independence when we estimate the proportions of potential outcome subgroups. The performances of POMI+IND and POMI+IND-R are very similar to each other, indicating that there is a minimal advantage in utilizing the information from the post-exposure variable $R$ to estimate these quantities in our simulations. Additional simulations with $N=500$ and $250$ show that the bias of the estimates becomes larger as the sample size decreases in most scenarios; however, they demonstrate similar patterns across different methods as N=1000 (See Tables 8-11 in Appendix).

\begin{table}
\begin{center}
\caption{Comparison of five methods in estimating the proportion of patients in each potential outcome subgroup. ESD: empirical standard deviation; SE: standard error; CR: coverage rate of 95$\%$ confidence interval. Here we assume potential outcomes conditional independence and no interaction between $X_1$ and $X_2$ (i.e., $\alpha=\beta_1=\beta_2=0$).}
\begin{tabular}{llccccc}
\hline
Quantities & Methods & True & 100 $\times$ Bias & ESD & SE & 100 $\times$ CR  \\
\hline
P(00) & POMI-Z & 0.21 & 0.15 & 2.781 & 4.911 & 98.6 \\
  & POMI & 0.21 & 0.40 & 2.947 & 4.819 & 97.6 \\
  & POMI+IND & 0.21 & -0.02 & 1.881 & 2.057 & 94.6 \\
  & POMI+IND-R & 0.21 & -0.05 & 1.891 & 2.058 & 96.0 \\
  & IPW & -- & -- & -- & -- & -- \\
\hline
 P(01) & POMI-Z & 0.10 & 1.01 & 2.449 & 4.667 & 98.4 \\
 & POMI & 0.10 & 0.71 & 2.454 & 4.615 & 98.2 \\
 & POMI+IND & 0.10 & 0.11 & 1.386 & 10..688 & 95.8 \\ & POMI+IND-R & 0.10 & 0.16 & 1.400 & 1.695 & 95.6 \\
& IPW & -- & -- & -- & -- & -- \\
 \hline
P(10) & POMI-Z & 0.29 & -0.85 & 2.980 & 5.046 & 98.6 \\
  & POMI & 0.29 & 0.45 & 3.171 & 4.968 & 97.0 \\
  & POMI+IND & 0.29 & 0.08 & 2.182 & 2.365 & 95.2 \\
  & POMI+IND-R & 0.29 & 0.16 & 2.174 & 2.395 & 95.6 \\
  & IPW & -- & -- & -- & -- & -- \\
\hline
P(11) & POMI-Z & 0.39 & -0.28 & 2.988 & 4.971 & 98.8 \\
& POMI & 0.39 & -0.69 & 2.908 & 4.934 & 98.0 \\
& POMI+IND & 0.39 & -0.17 & 2.153 & 2.263 & 94.4 \\
& POMI+IND-R & 0.39 & -0.26 & 2.189 & 2.324 & 94.8 \\
& IPW & -- & -- & -- & -- & -- \\
\hline
\end{tabular}
\end{center}
\label{Tab:2}
\end{table}

Third, we evaluate the performance of a generalized logit model for potential outcomes subgroup membership using POMI+IND (Table 4). The model is specified as $log[{\frac{P(ij)}{P(00)}}] = \gamma_{0ij} + \gamma_{1ij} X_2 + \gamma_{2ij} X_1$, where $ij = 01$, $10$, or $11$ and $\gamma_{0ij}, \gamma_{1ij}, \gamma_{2ij}$ are the corresponding coefficients. Since POMI+IND demonstrates advantages over POMI-Z and POMI and similarity as POMI+IND-R in performance based on prior simulations (Tables 2 and 3), we focus on the performance of POMI+IND in estimating the coefficients of the generalized logit model for the potential outcomes subgroup membership. Our simulations demonstrate that the POMI+IND method performs well in estimating the model coefficients with very little bias and their confidence intervals close to nominal $95\%$ coverage rates.

\begin{table}
\begin{center}
\caption{Properties of model coefficients from generalized logit model for potential outcomes subgroup membership using the POMI+IND method. ESD: empirical standard deviation; SE: standard error; CR: coverage rate of 95$\%$ confidence interval. Here we assume potential outcomes conditional independence and no interaction between $X_1$ and $X_2$ (i.e., $\alpha=\beta_1=\beta_2=0$).}
\begin{tabular}{llccccc}
\hline
Model & Parameter & True & 100 $\times$ Bias & ESD & SE & 100 $\times$ CR  \\
\hline
log[P(01)/P(00)] & $\gamma_{001}$ & -1.052 & -0.15 & 0.233 & 0.278 & 95.4 \\
  & $\gamma_{101}$ & 0.589 & -0.74 & 0.144 & 0.207 & 99.4 \\
  & $\gamma_{201}$ & 1.038 & 0.08 & 0.177 & 0.253 & 98.6 \\
\hline
log[P(10)/P(00)] & $\gamma_{010}$ & 0.002 & -0.15 & 0.176 & 0.208 & 96.8 \\
 & $\gamma_{110}$ & 0.498 & -2.31 & 0.174 & 0.197 & 96.0 \\
 & $\gamma_{210}$ & 0.996 & -1.70 & 0.200 & 0.244 & 95.6 \\
\hline
log[P(11)/P(00)] & $\gamma_{011}$ & -1.059 & -0.08 & 0.295 & 0.289 & 93.0 \\
 & $\gamma_{111}$ & 1.091 & -2.79 & 0.209 & 0.217 & 95.4 \\
 & $\gamma_{211}$ & 2.044 & -0.99 & 0.260 & 0.288 & 93.8 \\
\hline
\end{tabular}
\end{center}
\label{Tab:3}
\end{table}

Fourth, we continue to focus on the POMI+IND method and examine the impact of an unmeasured predictor $U$ of $Y$ on the quantities of interest (Figure 1). The presence of $U$ introduces a correlation between $Y_0$ and $Y_1$ and subsequently a violation of Assumption 4 (e.g. potential outcome conditional independence). We let $\beta_1=\beta_2$ and vary their sizes to be 0, 0.9, 1.06, 1.2, 1.38 and 1.49, which correspond to the proportions of variation of $Y$ explained by $U$ being $0\%, 5\%,10\%,15\%,20\%$ and $25\%$, respectively. As the correlations between $Y_0$ and $Y_1$ increase, the violation of Assumption 4 becomes more severe. Our simulations show that $U$ has little impact on the bias of ATE and MOR. However, as $U$ explains more of the variance in $Y$, $Y_0$ and $Y_1$ become increasingly more correlated; subsequently, the treatment-encouraged and treatment-discouraged subgroups become more rare and the bias of the estimated proportion in each potential outcome subgroup becomes more severe.

\begin{figure}
\centering
\includegraphics[width=13.5cm, height=11cm]{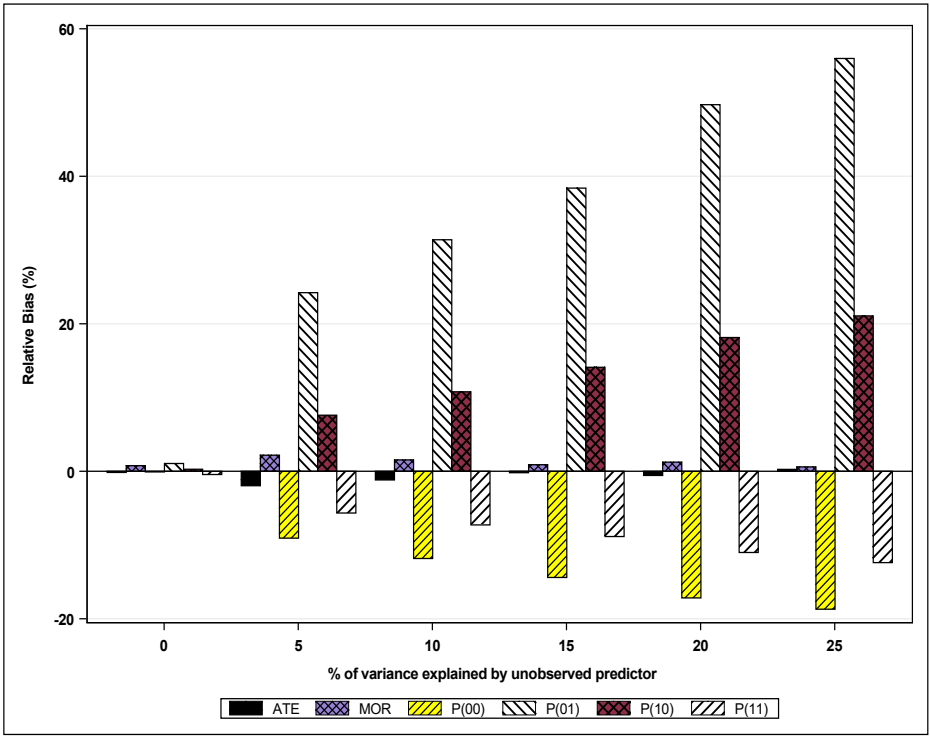}
  \caption{Relative bias of estimated average testing effect (ATE), marginal odds ratio (MOR), and proportion of patients in each potential outcomes subgroup, as the variance of $Y$ explained by an unmeasured predictor of $Y$, $U$, increases and subsequently violation of potential outcomes conditional independence increases (i.e., $\beta_1 \neq 0$ and $\beta_2 \neq 0$). Here we assume no interaction between $X_1$ and $X_2$ (i.e., $\alpha = 0$). The true values for ATE is -0.186, MOR is 0.458 and the proportions of patients in four potential outcomes subgroups, $P(00)$, $P(01)$, $P(10)$ and $P(11)$, are 0.212, 0.104, 0.290, and 0.394, respectively.}
   \label{fig:1}
\end{figure}
Lastly, we examine the robustness of the POMI+IND method towards model misspecification. We compare the method with the IPW method. For IPW to work properly, the model for the test assignment $Z$ needs to be correctly specified. Here, we specify the true model for $Z$ as $\textit{logit}[P(Z=1)] = - 1.5 - X_1 + 0.5 X_2+ 2 X_1 \times X_2$. However, in data analyses, neither IPW nor POMI+IND methods include the interaction term $X_1 \times X_2$. The simulation results are summarized in Table 5. It shows that IPW gives biased estimates of ATE and MOR and much below $95\%$ coverage rates for their confidence intervals; however, the model misspecification for $Z$ has little impact on estimates from the POMI+IND method. This highlights the need for IPW to specify correct models for $Z$; however, this is not required for the imputation method.

\begin{table}
\begin{center}
\caption{Comparing the robustness of POMI+IND and IPW towards model mis-specifications in estimating average testing effect (ATE) and marginal odds ratio (MOR). ESD: empirical standard deviation; SE: standard error; CR: coverage rate of 95$\%$ confidence interval. Here we assume potential outcomes conditional independence (i.e., $\beta_1=\beta_2=0$) and there is an interaction between $X_1$ and $X_2$ (i.e., $\alpha \neq 0$).}
\begin{tabular}{lllcccc}
\hline
Quantities & True & Methods & 100 $\times$ Bias & ESD & SE & 100 $\times$ CR  \\
\hline
ATE & -0.186 & IPW & -5.98 & 0.032 & 0.036 & 63.6 \\
  & -0.186 & POMI+IND & 0.12 & 0.032 & 0.033 & 95.0 \\
\hline
MOR & 0.459  & IPW &  -9.18 & 0.057 & 0.052 & 66.0 \\
 & 0.459 & POMI+IND & 0.65 & 0.065 & 0.063 & 94.0 \\
 \hline
\end{tabular}
\end{center}
\label{Tab:5}
\end{table}

\section{Application: the Impact of the RS Assay on Breast Cancer Chemotherapy Selection}
The analysis data set includes 2,931 incident breast cancer patients in Georgia and Los Angeles who were diagnosed between 2013-2014 with favorable prognosis (estrogen positive, human epidermal growth factor receptor 2 negative and invasive tumor behavior) \cite{Friese, Li}. These patients were chosen because the RS assay has the potential to make a difference in their chemotherapy treatment decisions. The data set comes from three data sources: 1) SEER databases for clinical features; 2) patient survey data from the ICanCare study; 3) the RS assay use and its testing results from the laboratory, Genomic Health, Inc. The outcome of interest is the receipt of chemotherapy (yes/no). The exposure of interest is the receipt of the RS assay (yes/no). The covariates include age (year), tumor grade (1, 2, or 3), tumor size ($\leq$1 cm, $1-2$ cm or $\geq$2 cm), node status (positive or negative), the number of major co-morbidities, menopausal status, and the risk of genetic mutation (low or high), education level (high school or less, some  college, graduate degree),  race (White, Black, Latino, Asian, other), insurance type (medicaid, medicare, private, other) and family income.

Overall, 35$\%$ of the patients received the RS assay and 27$\%$ received chemotherapy. Among these RS assay testers, the test results showed that $65\%$, $27\%$, and $8\%$ of patients received low-, intermediate- and high-risk scores respectively. We first fit a logistic regression with above covariates without addressing missing data. We find that the factors that are significantly associated with the use of chemotherapy include: the receipt of the RS assay (odds ratio (OR) = 0.42 with 95$\%$ confidence interval (CI): 0.32--0.55), age (OR=0.93, CI: 0.92--0.95), node status (positive vs. negative; OR=7.70, CI: 5.49 -- 10.80), tumor grade (2 vs. 1; OR=2.16, CI: 1.60 -- 2.93; 3 vs. 1; OR=14.42, CI: 9.82 -- 21.17), tumor size (1-2 vs. $\leq$ 1 cm; OR=2.41, CI: 1.72 -- 3.36; $\geq$ 2 vs.  $\leq$ 1 cm; OR=4.48, CI: 3.09 -- 6.48) and the risk of genetic mutation (high vs. low; OR=0.73, CI: 0.54 -- 0.99). The C-index from the logistic regression is 0.86. The model was conducted on 2,022 $(69\%)$ patients with complete data. (See Table 12 in Appendix).

We then apply the POMI+IND method to obtain causal estimates. We estimate that the average causal effect of the RS assay testing on chemotherapy (i.e., ATE) is $-10\%$ with its $95\%$ CI of $(-14\%,-7\%)$ and the MOR is 0.59 with its $95\%$ CI of $(0.49, 0.70)$. Table 6 presents the proportion of patients in each of the four potential outcome subgroups by overall and by covariates. Overall, it is estimated that the receipt of the RS assay encouraged a change of treatment plan in $9\% (8\%, 11\%)$ of patients from no chemotherapy to chemotherapy, and $20\% (17\%, 22\%)$ of patients from chemotherapy to no chemotherapy. Hence, the receipt of RS assay has reduced the overall chemotherapy use by $10\%$ and changed the treatment plan for a total of $30\%$ of patients. Regardless of whether they were tested or not, 59$\%$ of patients would never receive chemotherapy and 12$\%$ would always receive chemotherapy.

In addition, we find that, independent of the use of the RS assay, older patients ($\geq$65 years old) are more likely to reject chemotherapy than younger patients ($73\%$ vs. 49$\%$ estimated to be in the ``never-treatment'' subgroup). We also find that, with the use of the RS assay, patients were more likely to be discouraged from chemotherapy if they had a larger tumor ($36\%$ vs. $19\%$, $8\%$ in the ``treatment-discouraged'' subgroup for patients with tumor size of $\geq$ 2cm vs. $\leq$ 1cm, 1-2cm) or if they had positive nodes ($40\%$ vs. $15\%$ for patients with positive vs. negative nodes). It is particularly noteworthy that even though the guidelines did not make any recommendations about the use of the RS assay for node-positive patients, $14\%$ of them were offered the RS assay in practice. And in fact these patients were more likely to change their chemotherapy plans after testing with the RS assay than node-negative patients ($46\%$ vs. $25\%$ changed for patients with positive vs. negative nodes).

It is also evident that the RS assay played an influential role in chemotherapy decisions based on how much the results of the assay changed the treatment selections: for patients received a low-risk score, $2\%$ were encouraged to receive chemotherapy vs. $22\%$ discouraged from chemotherapy; for patients received a high-risk score, $34\%$ were encouraged vs. $5\%$ discouraged. In contrast, for patients received an intermediate-risk score, $19\%$ were encouraged vs. $17\%$ discouraged.

\begin{sidewaystable}
\begin{center}
\caption{Estimated proportion of patients in each of the four potential outcomes subgroups defined by the impact of the RS assay on chemotherapy according to patient characteristics, test results and insurance status.}
\begin{tabular}{llllll}
\hline
& &  \multicolumn{4}{c}{Estimated Proportion in Each} \\
& &  \multicolumn{4}{c}{Potential Outcomes Subgroup$_{(95\% \textup{CI})}$} \\
\hline
Variables & Categories & Never  & Treatment  & Treatment  & Always  \\
  &   &  Treated & Encouraged & Discouraged & Treated \\
\hline
Overall	& 		& 59$_{(56, 61)}$		&	9$_{(8, 11)}$	&	20$_{(17, 22)}$	&	12$_{(11, 14)}$	\\
\hline
Age	&	$<65$ years old	&	49$_{(46,51)}$	&	9$_{(8,11)}$	&	25$_{(23,28)}$	&	17$_{(15,19)}$	\\
	&	$\geq$65	&	73$_{(70,76)}$	&	10$_{(8,12)}$	&	12$_{(10,14)}$	&	6$_{( 4, 7)}$	\\
\hline
node	&	Negative	&	69$_{(66,71)}$	&	10$_{(9,12)}$	&	15$_{(13,16)}$	&	7$_{(6, 8)}$	\\
	&	Positive	&	20$_{(16,23)}$	&	6$_{(4, 8)}$ 	&	40$_{(34,46)}$	&	35$_{(29,41)}$	\\
\hline
Grade	&	1	&	79$_{(76,82)}$	&	6$_{( 4, 8)}$	&	13$_{(10,15)}$	&	2$_{( 1, 3)}$	\\
	&	2	&	56$_{(53,59)}$ 	&	10$_{(8,12)}$	&	24$_{(21,26)}$ 	&	10$_{( 8,12)}$	\\
	&	3	&	15$_{(11,20)}$	&	15$_{(11,19)}$	&	24$_{(19,29)}$ 	&	46$_{(40,51)}$	\\
\hline
Size	&	$<1$ cm	&	79$_{(76,83)}$	&	10$_{(8,12)}$	&	8$_{(6, 9)}$	&	3$_{(2, 5)}$	\\
	&	1-2 cm	&	59$_{(56,62)}$	&	10$_{(8,12)}$	&	19$_{(17,22)}$	&	12$_{(10,14)}$	\\
	&	$\geq$2 cm	&	31$_{(27,34)}$	&	8$_{(6,10)}$	&	36 $_{(31,42)}$	&	25$_{(21,30)}$	\\
\hline
21-gene score	&	Low-Risk Score	&	72$_{(70,75)}$	&	2$_{( 1, 3)}$	&	22$_{(20,25)}$	&	3$_{( 2, 5)}$	\\
	&	Intermediate-Risk Score 	&	42$_{(37,47)}$	&	19$_{(16,23)}$	&	17$_{(13,21)}$ 	&	21$_{(17,26)}$	\\
	&	High-Risk Score	&	5$_{( 1,10)}$	&	34$_{(25,43)}$	&	5 $_{( 1, 9)}$	&	56$_{(47,66)}$	\\
\hline
Insurance	&	Private	&	57$_{(35,79)}$	&	6$_{(0,18)}$	&	26$_{( 5,46)}$	&	11$_{(0,28)}$	\\
	&	Medicaid	&	48$_{(42,54)}$	&	10$_{( 6,14)}$	&	23$_{(18,28)}$	&	19$_{(14,25)}$	\\
	&	Medicare	&	75$_{(71,78)}$	&	10$_{( 7,12)}$	&	11$_{( 8,13)}$	&	5$_{(4, 7)}$	\\
\hline
\end{tabular}
\end{center}
\label{Tab:6}
\end{sidewaystable}

We also fit a multivariable multinomial logistic regression to examine the relationship between several key patient clinical predictors and the potential outcomes subgroup membership. These clinical predictors include tumor node status, tumor grade, tumor size, and age. They are the most important predictors for our study population and have been endorsed by the American Joint Committee on Cancer \cite{Kattan}. Table 7 listed the estimated probability of patients' potential outcome subgroup membership as a function of these clinical factors. We estimate that, for patients with negative node, tumor grade $1$, tumor size $\leq$ 1cm, and age$<$65 years old, there was a 7$\%$ chance to be encouraged to receive chemotherapy and 5$\%$ chance to be discouraged from chemotherapy by using the RS assay, an 88$\%$ chance of staying with the decision of no chemotherapy and a 1$\%$ of chance of staying with chemotherapy decision regardless of the use of the RS assay. We also estimate that, for patients with negative node, tumor grade of 3, tumor size of 1-2cm and age $<$ 65 years old, the chances of changing to chemotherapy from no chemotherapy, from no chemotherapy to chemotherapy or staying with chemotherapy or no chemotherapy after the use of the RS assay are 12$\%$, 28$\%$, $52\%$ and $8\%$ respectively. We also found that, generally, patients with a mix of low-risk factors (e.g., smaller tumor size, lower tumor grade, negative node) and high-risk factors (e.g., larger tumor size, higher tumor grade, positive node) for recurrence are more likely to be affected by using the RS assay than patients with either all low-risk or high-risk factors. For example, node-positive patients with small and low-grade tumors are most likely to be discouraged from chemotherapy by using the RS assay. Similarly, node-negative patients with small but high-grade tumor are most likely to be encouraged to take chemotherapy by using the RS assay. In contrast, patients with consistently low (high) risk clinical factors are more likely to remain in the ``never treated'' (``always treated'') subgroup.

\begin{table}
\begin{center}
\caption{Estimated probabilities of potential outcomes subgroup memberships according to patient clinical characteristics based on multinomial logistic regression}
\begin{tabular}{ccccccc}
\hline
&  & & \multicolumn{4}{c}{Estimated Probability in Each Subgroup$_{(95\% CI)}$ } \\
\hline
Tumor & Tumor & Age & Never & Treatment & Treatment & Always \\
Grade & Size & (yrs) & Treated & Encouraged & Discouraged & Treated \\
& (cm) & & & & & \\
\hline
\multicolumn{7}{c}{Node Status: Negative} \\
\hline
1	&	$\leq$ 1	&	$\leq 65$	&	0.88$_{(0.84, 0.92)}$	&	0.07$_{(0.04, 0.10)}$	&	0.05$_{(0.02, 0.07)}$	&	0.01$_{(0.00, 0.01)}$	\\
1	&	$\leq$ 1	&	$\geq 65$	&	0.93$_{(0.90, 0.96)}$	&	0.06$_{(0.03, 0.09)}$	&	0.01$_{(0.00, 0.02)}$	&	0.00$_{(0.00, 0.00)}$	\\
1	&	1-2	&	$\leq 65$	&	0.75$_{(0.70, 0.80)}$	&	0.06$_{(0.03, 0.09)}$	&	0.17$_{(0.13, 0.22)}$	&	0.02$_{(0.00, 0.03)}$	\\
1	&	1-2	&	$\geq 65$	&	0.92$_{(0.88, 0.95)}$	&	0.05$_{(0.03, 0.08)}$	&	0.03$_{(0.01, 0.05)}$	&	0.00$_{(0.00, 0.00)}$	\\
1	&	$\geq$ 2	&	$\leq 65$	&	0.54$_{(0.43, 0.65)}$	&	0.04$_{(0.00, 0.07)}$	&	0.37$_{(0.26, 0.47)}$	&	0.06$_{(0.01, 0.10)}$	\\
1	&	$\geq$ 2	&	$\geq 65$	&	0.87$_{(0.80, 0.94)}$	&	0.05$_{(0.00, 0.10)}$	&	0.07$_{(0.02, 0.12)}$	&	0.01$_{(0.00, 0.02)}$	\\
2	&	$\leq$ 1	&	$\leq 65$	&	0.75$_{(0.69, 0.80)}$	&	0.16$_{(0.11, 0.20)}$	&	0.08$_{ (0.04, 0.11)}$	&	0.02$_{(0.00, 0.04)}$	\\
2	&	$\leq$ 1	&	$\geq 65$	&	0.87$_{(0.82, 0.91)}$	&	0.10$_{(0.06, 0.14)}$	&	0.03$_{(0.01, 0.05)}$	&	0.00$_{(0.00, 0.01)}$	\\
2	&	1-2	&	$\leq 65$	&	0.54$_{(0.49, 0.60)}$	&	0.13$_{(0.09, 0.17)}$	&	0.25$_{(0.20, 0.30)}$	&	0.08$_{(0.05, 0.10)}$	\\
2	&	1-2	&	$\geq 65$	&	0.83$_{(0.79, 0.87)}$	&	0.09$_{(0.06, 0.12)}$	&	0.07$_{(0.04, 0.10)}$	&	0.01$_{(0.00, 0.01)}$	\\
2	&	$\geq$ 2	&	$\leq 65$	&	0.30$_{(0.23, 0.37)}$	&	0.10$_{(0.05, 0.14)}$	&	0.46$_{(0.38, 0.53)}$	&	0.15$_{(0.09, 0.20)}$	\\
2	&	$\geq$ 2	&	$\geq 65$	&	0.69$_{(0.61, 0.76)}$	&	0.12$_{(0.07, 0.17)}$	&	0.17$_{(0.11, 0.22)}$	&	0.03$_{(0.01, 0.05)}$	\\
3	&	$\leq$ 1	&	$\leq 65$	&	0.29$_{(0.15, 0.43)}$	&	0.30$_{(0.15, 0.45)}$	&	0.20$_{(0.07, 0.32)}$	&	0.21$_{(0.08, 0.35)}$	\\
3	&	$\leq$ 1	&	$\geq 65$	&	0.56$_{(0.41, 0.71)}$	&	0.30$_{(0.16, 0.44)}$	&	0.08$_{(0.02, 0.16)}$	&	0.05$_{(0.00, 0.11)}$	\\
3	&	1-2	&	$\leq 65$	&	0.08$_{(0.03, 0.12)}$	&	0.12$_{(0.07, 0.18)}$	&	0.28$_{(0.20, 0.36)}$	&	0.52$_{(0.44, 0.61)}$	\\
3	&	1-2	&	$\geq 65$	&	0.38$_{(0.27, 0.50)}$	&	0.26$_{(0.16, 0.36)}$ &	0.17$_{(0.09, 0.25)}$	&	0.18$_{(0.10, 0.27)}$	\\
3	&	$\geq$ 2	&	$\leq 65$	&	0.04$_{(0.01, 0.08)}$	&	0.07$_{(0.02, 0.12)}$	&	0.38$_{(0.27, 0.48)}$	&	0.51$_{(0.40, 0.62)}$	\\
3	&	$\geq$ 2	&	$\geq 65$	&	0.25$_{(0.12, 0.38)}$	&	0.21$_{(0.10, 0.33)}$	&	0.25$_{(0.13, 0.36)}$	&	0.29$_{(0.16, 0.42)}$	\\
\hline
\multicolumn{7}{c}{Node Status: Positive} \\
\hline
1	&	$\leq$ 1	&	$\leq 65$	&	0.40$_{(0.24, 0.55)}$	&	0.09$_{(0.01, 0.17)}$	&	0.39$_{(0.23, 0.55)}$	&	0.13$_{(0.02, 0.24)}$	\\
1	&	$\leq$ 1	&	$\geq 65$	&	0.65$_{(0.48, 0.82)}$	&	0.11$_{(0.01, 0.21)}$	&	0.20$_{(0.07, 0.33)}$	&	0.04$_{(0.00, 0.10)}$	\\
1	&	1-2	&	$\leq 65$	&	0.20$_{(0.11, 0.29)}$	&	0.06$_{(0.01, 0.11)}$	&	0.61$_{(0.49, 0.73)}$	&	0.13$_{(0.04, 0.21)}$	\\
1	&	1-2	&	$\geq 65$	&	0.56$_{(0.42, 0.71)}$	&	0.10$_{(0.02, 0.18)}$	&	0.29$_{(0.16, 0.43)}$	&	0.04$_{(0.00, 0.09)}$	\\
1	&	$\geq$ 2	&	$\leq 65$	&	0.11$_{(0.04, 0.18)}$	&	0.03$_{(0.00, 0.06)}$	&	0.63$_{(0.49, 0.77)}$	&	0.24$_{(0.11, 0.37)}$	\\
1	&	$\geq$ 2	&	$\geq 65$	&	0.44$_{(0.24, 0.63)}$	&	0.07$_{(0.00, 0.15)}$	&	0.37$_{(0.19, 0.55)}$	&	0.12$_{(0.01, 0.26)}$\\
2	&	$\leq$ 1	&	$\leq 65$	&	0.23$_{(0.10, 0.35)}$	&	0.09$_{(0.01, 0.17)}$	&	0.39$_{(0.23, 0.55)}$	&	0.30$_{(0.13, 0.46)}$	\\
2	&	$\leq$ 1	&	$\geq 65$	&	0.47$_{(0.29, 0.64)}$	&	0.09$_{(0.00, 0.17)}$	&	0.34$_{(0.17, 0.50)}$	&	0.11$_{(0.01, 0.22)}$	\\
2	&	1-2	&	$\leq 65$	&	0.09$_{(0.05, 0.14)}$	&	0.05$_{(0.02, 0.08)}$	&	0.51$_{(0.42, 0.60)}$	&	0.35$_{(0.26, 0.43)}$	\\
2	&	1-2	&	$\geq 65$	&	0.34$_{(0.23, 0.45)}$	&	0.07$_{(0.02, 0.13)}$	&	0.45$_{(0.34, 0.57)}$	&	0.13$_{(0.05, 0.20)}$	\\
2	&	$\geq$ 2	&	$\leq 65$	&	0.05$_{(0.02, 0.07)}$ &	0.03$_{(0.00, 0.05)}$	&	0.52$_{(0.43, 0.61)}$	&	0.40$_{(0.31, 0.50)}$\\
2	&	$\geq$ 2	&	$\geq 65$	&	0.21$_{(0.12, 0.30)}$	&	0.07$_{(0.02, 0.12)}$	&	0.49$_{(0.37, 0.60)}$ &	0.24$_{(0.14, 0.33)}$	\\
3	&	$\leq$ 1	&	$\leq 65$	&	0.03$_{(0.00, 0.09)}$	&	0.11$_{(0.00, 0.25)}$&	0.17$_{(0.01, 0.35)}$	&	0.69$_{(0.43, 0.94)}$	\\
3	&	$\leq$ 1	&	$\geq 65$	&	0.13$_{(0.00, 0.31)}$	& 0.18$_{(0.00, 0.40)}$	&	0.18$_{(0.01, 0.38)}$	&	0.50$_{(0.19, 0.82)}$	\\
3	&	1-2	&	$\leq 65$	&	0.01$_{(0.00, 0.02)}$	&	0.05$_{(0.00, 0.10)}$	&	0.16$_{(0.06, 0.26)}$	&	0.79$_{(0.67, 0.90)}$	\\
3	&	1-2	&	$\geq 65$	&	0.06$_{(0.00, 0.14)}$	&	0.13$_{(0.01, 0.25)}$	&	0.19$_{(0.06, 0.32)}$	&	0.61$_{(0.43, 0.80)}$	\\
3	&	$\geq$ 2	&	$\leq 65$	&	0.01$_{(0.00, 0.02)}$	&	0.04$_{(0.00, 0.07)}$	&	0.19$_{(0.10, 0.28)}$	&	0.77$_{(0.67, 0.86)}$	\\
3	&	$\geq$ 2	&	$\geq 65$	&	0.04$_{(0.00, 0.09)}$	&	0.10$_{(0.01, 0.19)}$	&	0.18$_{(0.07, 0.28)}$	&	0.68$_{(0.53, 0.83)}$	\\
 \hline
\end{tabular}
\end{center}
\label{table x}
\end{table}

We further examine how robust our analysis results are towards the violation of the potential outcomes conditional independence assumption (Assumption 4). Since $Y_0$ and $Y_1$ are not observed simultaneously in our data, we cannot evaluate this conditional independence assumption nor estimate the correlation between $Y_0$ and $Y_1$ based on our data. We artificially induce a correlation by simulating an additional predictor U of Y that is not in the data set, and examine how sensitive our analyses are towards the correlation between $Y_0$ and $Y_1$. Since $U$ is a predictor of $Y$, $U$ is a common predictor of both $Y_0$ and $Y_1$. Since U is not observed or adjusted for, this induces the correlation between $Y_0$ and $Y_1$. The more correlated $Y_0$ and $Y_1$ are, the more severe the violation of the conditional independence assumption is. We let $U = (-0.267 + Y)*b_1 + e$, where $Y$ is the observed receipt of chemotherapy and $e \sim$ Normal$(0, \sigma^2)$. Since $26.7\%$ of patients were observed to receive chemotherapy, we choose -0.267 in this equation to ensure that the mean of $U$ equals $0$. We vary $b_1$ to change the association between $U$ and $Y$ and examine how it affects the causal estimates. For each $b_1$ value, we re-impute the missing values conditional on the same set of covariates listed previously and the simulated predictor $U$, and then re-estimate the causal quantities of interest using the POMI+IND method. Figure 2 plots the estimates according to a wide range of the extent of variation in $Y$ explained by $U$, which is estimated by a non-linear mixed effect model with $U$ as a random effect \cite{Browne}. As the extent of variation in $Y$ explained by $U$ increases, the proportions of patients in the ``never treated'' and ``always treated'' subgroups increase, and the proportions in other two subgroups decrease. Subsequently, the overall influence of 21-gene assay on chemotherapy receipts decreases. We also observe that MOR and ATE only increase slightly. Overall, these causal estimates are not very sensitive towards $U$.

\begin{figure}[!tbp]
  \centering
  \begin{minipage}[b]{0.495\textwidth}
    \includegraphics[width=\textwidth]{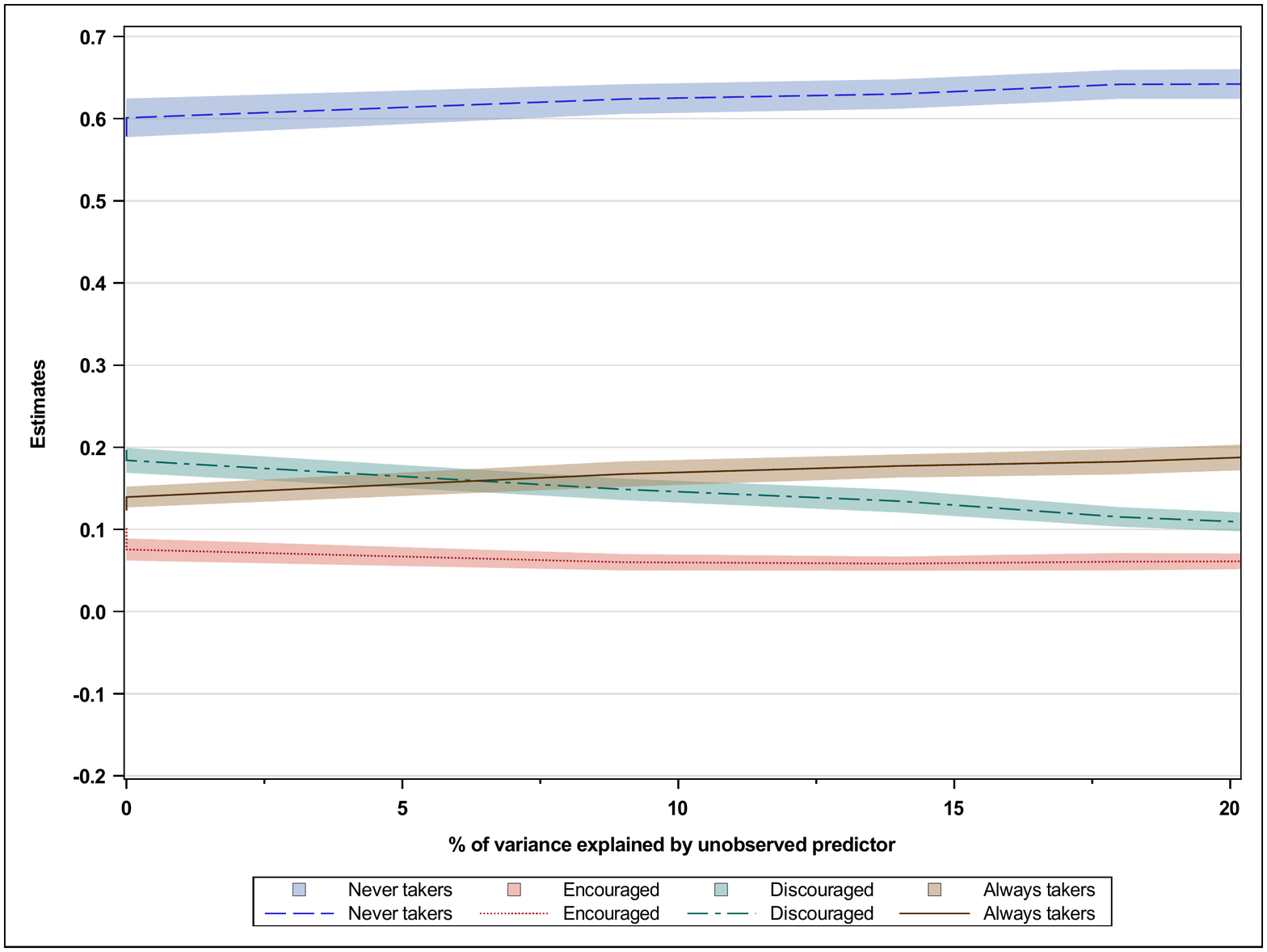}
  \end{minipage}
  \hfill
  \begin{minipage}[b]{0.495\textwidth}
    \includegraphics[width=\textwidth]{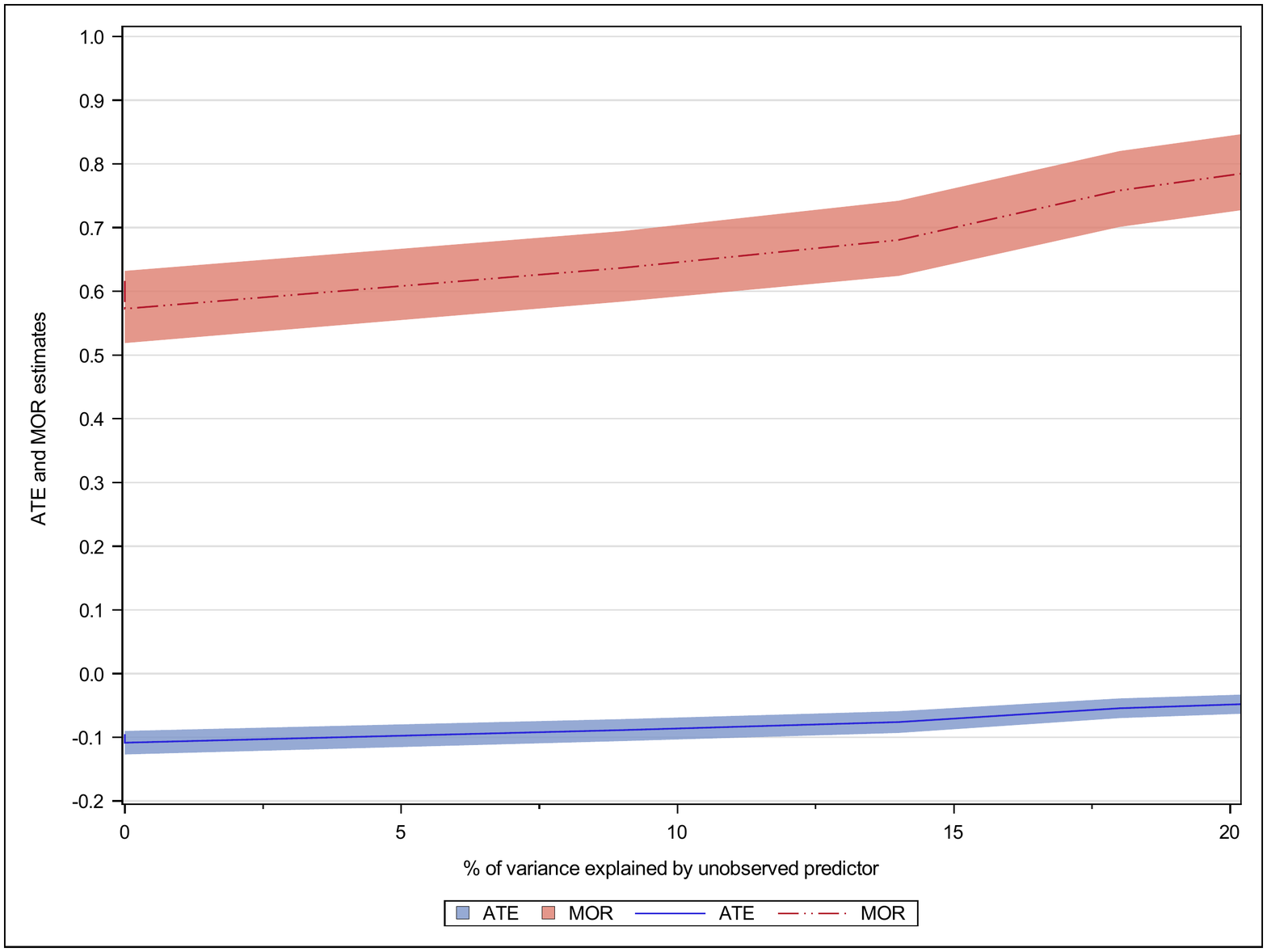}
  \end{minipage}
  \caption{Estimated proportion of patients in each potential outcomes subgroup, ATE and MOR, as the variance of $Y$ explained by an unmeasured predictor of $Y$, $U$, increases and the subsequent violation of the potential outcomes conditional independence assumption increases. }
  \end{figure}

\section{Discussion}

We proposed a potential outcomes framework to measure the influence of medical testing on treatment selection, stratifying patients into four potential outcomes subgroups based on whether the test has any impact on treatments election (either undergoing the treatment or not), and whether the test encourages or discourages treatment selection. This framework captures well the differential causal effect of testing on treatment selection for different patient subgroups, and allows us to examine the distribution of these subgroups and characteristics associated with these subgroup memberships. The classification suggests targets for more efficient utilization of medical tests. With advanced tests designed to support precision medicine becoming increasingly available and increasing concern for over-testing \cite{Schnipper}, the causal estimates produced from this framework become more relevant. The framework also allows us to estimate many traditional causal quantities of interest (e.g., average causal effects, marginal odds ratios) on which existing causal inference methods more often focus. By taking advantage of the connection between missing data and causal inference, we can simultaneously implement the causal framework as well as handle missing data in the observable variables (i.e., due to non-response). We also explore the advantages of incorporating unique features of causal inference assumptions in the data imputation process. These assumptions are often made in causal inference literature, but cannot be estimated or verified from the data. Simulation studies suggested that the proposed methods provide reasonably accurate inference. Additionally, the explicit incorporation of the causal assumptions in the data imputation process can improve precision for some causal estimates. We also find that bias can occur for some causal estimates when the potential outcomes conditional independence assumption is violated, which can be assessed through sensitivity analyses. The proposed framework was then used to examine how the most commonly used genomic test in the United States, 21-gene assay, impacts the use of chemotherapy among breast cancer patients.

Existing research has often focused on estimating the overall relationship between medical testing and treatment options in a population. For example, breast magnetic resonance imaging has been found to be associated with receiving a mastectomy instead of breast conserving surgery (adjusted odds ratio = 1.21) \cite{Killelea}. In another example, it is found that $55\%$ of women with invasive breast cancer and positive HER2 genetic testing results received trastuzumab treatment \cite{Goddard}. Although this information is of interest, the average effect of testing on treatment choice may not apply equally across subgroups of patients who may not benefit from (or, may be harmed by) testing. Hence, it is important to recognize that a medical test can have different influences on treatment decisions for different patients; i.e., there may be effect heterogeneity of testing on treatment. We proposed the use of potential outcomes subgroups to classify patients by the presence (vs. absence) and direction of the effect of testing on treatment decisions. Patients are classified into one of the four groups: 1) never treated; 2) treatment encouraged; 3) treatment discouraged; 4) always treated. These subgroups naturally capture the effect heterogeneity of medical testing on treatment decisions. Additionally, we investigated how patients' membership in these subgroups may be driven by patient characteristics or systemic factors. Such classification would suggest targets for improving the utilization of medical tests and the allocation of limited resources. For example, in the data analyses, we found that patients with positive nodes were more influenced by the testing and more of these patients were discouraged from receiving chemotherapy than patients with negative nodes, even though guidelines make no recommendations for how to use the 21-gene assay in node-positive patients. We also find that patients with a mix of low and high risk factors were more likely to be influenced by the use of 21-gene assay, and the tests were more likely to either encourage or discourage these patients for chemotherapy use. Given that the test was recommended to aid decision-making in people with intermediate risk of recurrence, our results show that efforts to ensure that the patients with a mix of low and high risk factors are offered testing are likely to produce substantial changes in plans and maximize the utility of the test.

Our methods differ from traditional latent class models, which usually deal with multiple or longitudinal outcomes and do not classify patients by potential outcomes. Subgroup analysis or adding interaction terms can also investigate effect heterogeneity; however, these approaches require us to pre-specify variables that modify effect and suffer from false discovery rate \cite{Fink}. In summary, standard analyses cannot provide the classification targeted by our methods. Our methods of analyzing heterogeneous testing effect using the potential outcomes subgroups approach are at the individual level. For each patient, the testing effect on treatment is obtained by comparing her underlying tendency of receiving treatment with vs. without a given test. Our approach may be complementary to the standard analyses based on known effect modifiers. Further discussions of individual heterogeneous treatment effects are discussed by others \cite{Zhang, Poulson, Egleston}.

We proposed the use of multiple imputation methods to impute the missing potential outcomes for each patient ($Y_0$ or $Y_1$) since we do not observe two potential outcomes of the same patient simultaneously. Subsequently, we can classify patients into four potential outcomes subgroups according to individual heterogeneous treatment effects. We chose the sequential regression multiple imputation method. It has the advantages of accommodating many different data types and structure compared with other missing data methods \cite{Ambler}. The main uncertainty in theoretical justification for this method is the likely incompatibility between the assumed conditional models and the true conditional models derived from the joint conditional distributions \cite{Raghu}. However, many simulation studies have demonstrated the robustness of this method towards possible misspecifications of the assumed conditional models \cite{Ambler}.

Our methods have the advantage of imputing missing potential outcomes and regular missing values in the observable variables simultaneously during the data imputation process. This is more convenient and faster than most of other causal inference methods in the presence of missing data, which will likely need to handle the missing data problem first using multiple imputation methods before applying causal inference methods and combining analysis results across multiple imputed data sets.

The use of multiple imputation methods utilizes the close connection between causal inference and missing data framework, since the causal inference problems here are essentially a missing data problem. However, there are some unique features of causal inference that are different from the regular missing data problems in our context. First, causal assumptions are unique. Even though there is no information from the data to test the validity of the assumptions regarding ignorability or potential outcomes conditional independence, we demonstrate that the explicit incorporation of these causal assumptions into the data imputation process can increase the precision of some of our causal estimates such as the proportion of patients in each of the potential outcomes subgroups. Second, we find that the violation of the potential outcomes conditional independence assumption does not have much impact on traditional causal quantities such as the average causal effect or marginal odds ratio. However, it does impact the causal quantities regarding the potential outcomes subgroups, which concern the joint distribution of the potential outcomes. As the correlation between two potential outcomes increases, patients in the treatment-encouraged and treatment-discouraged subgroups become fewer. Since there is no information from the empirical data about the validity of this assumption, we proposed a method to conduct sensitivity analyses to relax this assumption and evaluate the impact of violation of this assumption. Our proposed sensitivity analyses are closely related to the random effect models proposed by Zhang et al. \cite{Zhang}, which use random effects to induce the correlation between $Y_0$ and $Y_1$. Here we used an unmeasured predictor of $Y$ to induce the correlation. Third, $Z$ is the missing data indicator for $Y_0$ and $Y_1$. With an MAR assumption, multiple imputation does not usually need to include the missing indicator \cite{Van}. However, $Z$ is also the exposure of interest. And it is a function of the covariate $X_2$, which is partially missing. Hence, for multiple imputation of $X_2$ to work properly, the imputation model for $X_2$ needs to include $Z$ to satisfy the MAR assumption. This explains why POMI-Z performs more poorly in our simulations compared with methods including $Z$ in the imputation models. Fourth, a post-exposure variable (e.g., testing results) that is correlated with the outcome is not usually utilized in the causal inference methods because it can be in the causal pathway between the exposure and outcome. However, it is usually used in the multiple imputation process and not treated differently from other variables. Our study finds that the incorporation of a post-exposure variable in the multiple imputation process still generates valid causal estimates. In fact, it did not make much difference whether or not we included the post-exposure variable in our simulations. However, we believe that if the post-exposure variable is correlated with variables that have more missingness, the utilization of the post-exposure variable will increase precision. In our simulations, there is only $10\%$ missingness in the variable $Y$, a correlate of the post-exposure variable $R$; hence, the incorporation of $R$ in the imputation process did not provide noticeable efficiency gain. Lastly, we find that our POMI methods do not require the need to correctly specify the model for the exposure to obtain valid causal estimates; however, a propensity-score based causal method usually does. This feature occurs because the POMI methods do not require one to model the exposure but only to provide a correct predictive model for the outcome. Our methods can be implemented through IVEware in SAS, or through the MICE (Multivariate Imputation by Chained Equations) package in R. We have provided code examples in GitHub at https://github.com/yunliyunli/POMI.

In summary, we proposed a potential outcomes framework to classify patients according to the differential causal effect of medical testing on treatment selection for different individual patients. It has the potential to be very useful in examining effect heterogeneity of medical testing on treatment selection and in proposing targets for the optimal use of medical tests. Future research includes conducting more sensitivity analyses on the causal assumptions and improving the efficiency for these causal quantities of interest.

\begin{dci}
The authors declared no potential conflicts of interest with respect to the research, authorship,
and/or publication of this article.
\end{dci}

\begin{funding}
This work was supported by the National Cancer Institute through grants P01CA163233, CA46592, CA129102 and U057141.
\end{funding}

\begin{acks}
Yun Li and Irina Bondarenko contribute equally.
\end{acks}
  
\begin{spacing}{1}	

\end{spacing}

\section{Appendix}

\section{Additional Simulation Results with Different Sample Sizes}

Additional simulations were conducted using a sample size of 250 to evaluate the impact of sample size. Results are summarized in Tables 8-9. 
\begin{table}
\begin{center}
\caption{Additional simulations: comparison of five methods in estimating average testing effect (ATE), and marginal odds ratios (MOR) when N=250. ESD: empirical standard deviation; SE: average standard error; CR: coverage rate of 95$\%$ confidence interval. Here we assume potential outcomes conditional independence and no interaction between $X_1$ and $X_2$ (i.e., $\alpha=\beta_1=\beta_2=0$).}
\begin{tabular}{llccccl}
\hline
Quantities & Methods & True & 100 $\times$ Bias & ESD & SE & 100 $\times$ CR \\
\hline
ATE 	&	POMI-Z	&	-0.186	&	2.9	&	0.071	&	0.076	&	93.2	\\
	&	POMI 	&	-0.186	&	1.6	&	0.070	&	0.082	&	94	\\
	&	POMI+IND	&	-0.186	&	0.8	&	0.068	&	0.066	&	91.4	\\
	&	POMI+IND-R 	&	-0.186	&	0.6	&	0.066	&	0.066	&	92.4	\\
	&	IPW 	&	-0.186	&	-0.5	&	0.074	&	0.077	&	93.8	\\
\hline
MOR 	&	POMI-Z	&	0.458	&	7.7	&	0.169	&	0.187	&	91.8	\\
	&	POMI 	&	0.458	&	4.7	&	0.161	&	0.192	&	95.6	\\
	&	POMI+IND	&	0.458	&	3.1	&	0.147	&	0.141	&	90.6	\\
	&	POMI+IND-R 	&	0.458	&	2.4	&	0.143	&	0.141	&	92.8	\\
	&	IPW 	&	0.458	&	2.1	&	0.146	&	0.156	&	93.6	\\
\hline
\end{tabular}
\end{center}
\label{Tab:8}
\end{table}
 
\begin{table}
\begin{center}
\caption{Additional simulation: comparison of five methods in estimating the proportion of patients in each potential outcome subgroup when N=250. ESD: empirical standard deviation; SE: standard error; CR: coverage rate of 95$\%$ confidence interval. Here we assume potential outcomes conditional independence and no interaction between $X_1$ and $X_2$ (i.e., $\alpha=\beta_1=\beta_2=0$).}
\begin{tabular}{llccccc}
\hline
Quantities & Methods & True & 100 $\times$ Bias & ESD & SE & 100 $\times$ CR  \\
\hline
P(00)	&	POMI-Z	&	0.21	&	0.49	&	5.527	&	7.529	&	94.4	\\
	&	POMI 	&	0.21	&	0.44	&	4.703	&	7.934	&	98.2	\\
	&	POMI+IND	&	0.21	&	-0.27	&	3.576	&	4.046	&	95.4	\\
	&	POMI+IND-R 	&	0.21	&	-0.27	&	3.654	&	4.063	&	95.2	\\
    & IPW & -- & -- & -- & -- & -- \\
\hline
P(01)	&	POMI-Z	&	0.10	&	3.08	&	4.737	&	6.978	&	91.6	\\
	&	POMI 	&	0.10	&	2.82	&	3.965	&	7.391	&	98	\\
	&	POMI+IND	&	0.10	&	0.60	&	2.933	&	3.368	&	95	\\
	&	POMI+IND-R 	&	0.10	&	0.51	&	2.839	&	3.383	&	95.8	\\
    & IPW & -- & -- & -- & -- & -- \\
 \hline
P(10)	&	POMI-Z	&	0.29	&	-1.11	&	6.328	&	8.268	&	94	\\
	&	POMI 	&	0.29	&	-0.45	&	5.724	&	8.988	&	98.4	\\
	&	POMI+IND	&	0.29	&	-0.24	&	4.511	&	4.700	&	93.2	\\
	&	POMI+IND-R 	&	0.29	&	-0.04	&	4.438	&	4.712	&	94.2	\\
  & IPW & -- & -- & -- & -- & -- \\
\hline
P(11)	&	POMI-Z	&	0.39	&	-0.43	&	6.261	&	8.080	&	95.4	\\
	&	POMI 	&	0.39	&	-0.39	&	5.211	&	8.865	&	99.4	\\
	&	POMI+IND	&	0.39	&	-0.08	&	4.073	&	4.689	&	96.8	\\
	&	POMI+IND-R 	&	0.39	&	-0.20	&	4.001	&	4.670	&	96.2	\\
    & IPW & -- & -- & -- & -- & -- \\
\hline
\end{tabular}
\end{center}
\label{Tab:9}
\end{table}

Additional simulations were conducted using a sample size of 500 to evaluate the impact of sample size. Results are summarized in Tables 10-11. 

\begin{table}
\begin{center}
\caption{Additional simulation: comparison of five methods in estimating average testing effect (ATE), and marginal odds ratios (MOR) when N=500. ESD: empirical standard deviation; SE: average standard error; CR: coverage rate of 95$\%$ confidence interval. Here we assume potential outcomes conditional independence and no interaction between $X_1$ and $X_2$ (i.e., $\alpha=\beta_1=\beta_2=0$).}
\begin{tabular}{llccccl}
\hline
Quantities & Methods & True & 100 $\times$ Bias & ESD & SE & 100 $\times$ CR \\
\hline
ATE 	&	POMI-Z	&	-0.186	&	2.0	&	0.048	&	0.047	&	90.6	\\
	&	POMI 	&	-0.186	&	0.4	&	0.047	&	0.050	&	93.8	\\
	&	POMI+IND	&	-0.186	&	0.1	&	0.048	&	0.047	&	91.6	\\
	&	POMI+IND-R 	&	-0.186	&	0.1	&	0.049	&	0.047	&	92	\\
	&	IPW 	&	-0.186	&	-0.9	&	0.052	&	0.054	&	96	\\
MOR 	&	POMI-Z	&	0.458	&	4.8	&	0.110	&	0.105	&	90.2	\\
	&	POMI 	&	0.458	&	1.6	&	0.099	&	0.104	&	94.2	\\
	&	POMI+IND	&	0.458	&	0.9	&	0.101	&	0.096	&	91.6	\\
	&	POMI+IND-R 	&	0.458	&	0.8	&	0.100	&	0.096	&	92.6	\\
	&	IPW 	&	0.458	&	0.4	&	0.101	&	0.106	&	95.2	\\
\hline
\end{tabular}
\end{center}
\label{Tab:10}
\end{table}

\begin{table}
\begin{center}
\caption{Additional simulation: comparison of five methods in estimating the proportion of patients in each potential outcome subgroup when N=500. ESD: empirical standard deviation; SE: standard error; CR: coverage rate of 95$\%$ confidence interval. Here we assume potential outcomes conditional independence and no interaction between $X_1$ and $X_2$ (i.e., $\alpha=\beta_1=\beta_2=0$).}
\begin{tabular}{llccccc}
\hline
Quantities & Methods & True & 100 $\times$ Bias & ESD & SE & 100 $\times$ CR  \\
\hline
P(00)	&	POMI-Z	&	0.21	&	0.34	&	4.315	&	5.948	&	96	\\
	&	POMI 	&	0.21	&	0.03	&	3.737	&	6.154	&	97.4	\\
	&	POMI+IND	&	0.21	&	-0.07	&	2.734	&	2.863	&	93.8	\\
	&	POMI+IND-R 	&	0.21	&	-0.07	&	2.777	&	2.925	&	93.2	\\
    & IPW & -- & -- & -- & -- & -- \\
\hline
P(01)	&	POMI-Z	&	0.10	&	1.50	&	1.501	&	0.954	&	95.4	\\
	&	POMI 	&	0.10	&	0.96	&	0.963	&	0.976	&	97.6	\\
	&	POMI+IND	&	0.10	&	0.23	&	0.228	&	0.954	&	95.4	\\
	&	POMI+IND-R 	&	0.10	&	0.29	&	0.291	&	0.962	&	96.2	\\
    & IPW & -- & -- & -- & -- & -- \\
 \hline
P(10)	&	POMI-Z	&	0.29	&	-0.84	&	4.522	&	6.278	&	95.2	\\
	&	POMI 	&	0.29	&	0.11	&	4.167	&	6.523	&	98.6	\\
	&	POMI+IND	&	0.29	&	0.14	&	3.289	&	3.357	&	93.2	\\
	&	POMI+IND-R 	&	0.29	&	0.23	&	3.303	&	3.387	&	93.4	\\
  & IPW & -- & -- & -- & -- & -- \\
\hline
P(11)	&	POMI-Z	&	0.39	&	-0.47	&	4.231	&	6.325	&	98	\\
	&	POMI 	&	0.39	&	-0.59	&	3.934	&	6.312	&	99	\\
	&	POMI+IND	&	0.39	&	-0.30	&	2.928	&	3.296	&	95.6	\\
	&	POMI+IND-R 	&	0.39	&	-0.45	&	2.911	&	3.302	&	96	\\
    & IPW & -- & -- & -- & -- & -- \\
\hline
\end{tabular}
\end{center}
\label{Tab:11}
\end{table}

\section{Logistic regression Results for Chemotherapy Use}

 The results from the logistic regression model are presented in Table 12.

\begin{table}
\begin{center}
\caption{Logistic Regression Results for Chemotherapy Use Based on Observed Variables. RS Assay: 21-Gene Assay}
\begin{tabular}{llccc}
 \hline
Variable & Categories & OR & \multicolumn{2}{c}{$95\%$ CI} \\
\hline
Receipt of RS Assay  & Yes vs. No &  0.42 & 0.32 & 0.55 \\
Tumor grade & 2	&	2.16	&	1.60	&	2.93	\\
 \ \ \ \ \ \ \ \ (Ref: 1)   & 3	&	14.42	&	9.82	&	21.17	\\
Age (year) &	&	0.93	&	0.92	&	0.95	\\
Tumor size (cm) & $1-2$	&	2.41	&	1.72	&	3.36	\\
\ \ \ \ \ \ \ \ (Ref: $\leq$1)  & $\geq$ 2 &	4.48	&	3.09	&	6.48	\\
Positive Node Status & Yes vs. No	&	7.70	&	5.49	&	10.80	\\
High Risk of Genetic Mutation & Yes vs. No & 0.73 & 0.54 & 0.99 \\
Number of Comorbidities & & 1.04 & 0.84 & 1.28 \\
Race & Black & 1.05 & 0.73 & 1.51 \\
\ \ \ \ \ \ \ \ (Ref: White) & Latino & 0.73 & 0.50 & 1.07 \\
 & Asian & 0.77 & 0.47 & 1.24 \\
 & Other & 1.12 & 0.35 & 3.58 \\
Education &                Some college	&	0.87	&	0.60	&	1.26	\\
\ \ \ \ \ \ \ \ (Ref: High school or less)            &         Graduate	&	0.90	&	0.61	&	1.32	\\
Family Income & & 1.02 & 0.94 & 1.11 \\
Insurance        &     None	&	0.78	&	0.19	&	3.25	\\
\ \ \ \ \ \ \ \ (Ref: Private)        &     Medicaid 	&	1.39	&	0.87	&	2.23	\\
         &    Medicare &	0.92	&	0.62	&	1.36	\\
        &     Other	&	1.56	&	0.44	&	5.51	\\
\hline
\end{tabular}
\end{center}
\label{table 12}
\end{table}

\section{Variance Estimates of Marginal Odds Ratio}
The MOR is defined as MOR= $[P_1/(1-P_1)] /[P_0/(1-P_0)]$. Since MOR is skewed and bounded to be positive, we use log transformation to make it more symmetric so that we can use the delta method to obtain the variance estimate of $log(\widehat{MOR})$ first. Let  $log(\widehat{MOR})_d$ be the point estimate of log(MOR) in the $d$th imputed data set where $d=1, \ldots\, D$. The combined multiple imputation estimate of log(MOR) across D imputed data sets is
\begin{eqnarray}
\overline{log(\widehat{MOR})} = \frac{1}{D} \sum_{d=1}^D log(\widehat{MOR})_d. \nonumber
\end{eqnarray}
For the variance of $log(\widehat{MOR})$ in the dth imputed data set denoted by $W_d$, we calculate it using the delta method as below:
\begin{eqnarray}
&& \begin{bmatrix}
\frac{\partial log(\widehat{MOR})}{\partial p_1} & \frac{\partial log(\widehat{MOR})}{\partial p_0}
\end{bmatrix} \times Var(\widehat{p}_1, \widehat{p}_0) \times \begin{bmatrix}
\frac{\partial log(\widehat{MOR})}{\partial p_1}  \\
\frac{\partial log(\widehat{MOR})}{\partial p_0}
\end{bmatrix} \nonumber \\
&&= \frac{1}{N}  \begin{bmatrix}
\frac{1}{\widehat{p}_1(1-\widehat{p}_1)} & \frac{-1}{\widehat{p}_0(1-\widehat{p}_0)}
\end{bmatrix} \times \begin{bmatrix}\widehat{p}_1(1-\widehat{p}_1) & \widehat{p}_{11} - \widehat{p}_0\widehat{p}_1 \\ \widehat{p}_{11} - \widehat{p}_0\widehat{p}_1 & \widehat{p}_0(1 - \widehat{p}_0) \end{bmatrix} \times \begin{bmatrix}
\frac{1}{\widehat{p}_1(1-\widehat{p}_1)} \\ \frac{-1}{\widehat{p}_0(1-\widehat{p}_0)}
\end{bmatrix} \nonumber \\
&&= \frac{\widehat{p}_0 + \widehat{p}_1 - (\widehat{p}_0 -\widehat{p}_1)^2 - 2\widehat{p}_{11}}{N\widehat{p}_0\widehat{p}_1(1-\widehat{p}_0)(1 - \widehat{p}_1)} \label{within}
\end{eqnarray}

Averaging over multiple imputed data sets, we calculate the average within-imputation variance as $\overline{W} = \sum_{d=1}^D W_d/D$. For between-imputation variance denoted by $B$, we have $B = \sum_{d=1}^D \left[log(\widehat{MOR})_d - \overline{log(\widehat{MOR})}\right]^2 / (D-1)$. The multiple imputation estimate of the variance of $\overline{log(\widehat{MOR})}$ is $V = \overline{W} + (1 + \frac{1}{D})B$.

After we obtain the variance of the estimated log(MOR), we can then use the delta method to obtain the variance of estimated MOR such that $Var(\widehat{MOR}) = \exp\left\{2\left[log(\widehat{MOR})\right]\right\}\times Var\left[log(\widehat{MOR})\right]$.

\end{document}